\documentclass[11pt]{article}

\usepackage{authblk}

\usepackage[english]{babel}															% English language/hyphenation
\usepackage{amsmath}										% Math packages
\usepackage{amsfonts}
\usepackage{amsthm}
\usepackage{latexsym}

\usepackage{graphicx}
\usepackage{epstopdf}

\usepackage{url}
\usepackage{amssymb}
\usepackage{bbm}
\usepackage{MnSymbol}

\usepackage{eucal}
\usepackage{mathrsfs}
\usepackage{amscd}

\usepackage{sectsty}												% Custom sectioning (see below)
\allsectionsfont{\centering \normalfont\scshape}	% Change font of al section commands

\usepackage{mathtools}

\makeatletter
\numberwithin{equation}{section}

\newtheorem{theorem}{Theorem}

\title{
		\vspace{-1in} 	
		\usefont{OT1}{bch}{b}{n}
		\normalfont \normalsize \textsc{} \\ [25pt]
		\huge  Explicit formulas for reaction  probability  in reaction-diffusion experiments
}

 \author[1]{M. Wallace\thanks{One Brookings Dr., St. Louis, MO, 63130; matt@math.wustl.edu}}
\author[1]{R. Feres\thanks{One Brookings Dr., St. Louis, MO, 63130; feres@math.wustl.edu}}
\author[2]{G. Yablonsky\thanks{3450 Lindell Blvd, St. Louis, MO, 63103; gyablons@slu.edu}}
\author[1]{A. Stern\thanks{One Brookings Dr., St. Louis, MO, 63130; astern@math.wustl.edu}}
\affil[1]{\normalsize Department of Mathematics, Washington University}
\affil[2]{\normalsize Parks College of Engineering, Aviation and Technology, Saint Louis University}

\begin{document}

\date{}

\maketitle

\begin{abstract}
A computational procedure is developed for determining  the conversion probability for
 reaction-diffusion systems in which a first-order catalytic reaction is performed over active particles.
 We apply this general method to systems on metric graphs, which may be viewed as $1$-dimensional approximations of
 $3$-dimensional systems,  and obtain explicit formulas for conversion. We then
 study numerically a class of $3$-dimensional systems and test how accurately  they are described by model formulas obtained for metric
 graphs.  The optimal arrangement  of active particles in a
 $1$-dimensional multiparticle system  is found, which is shown to depend on the level of catalytic activity:  conversion
 is maximized for low catalytic activity when all particles are bunched together close to the point of gas injection, and
 for high catalytic activity when the particles are evenly spaced.
\end{abstract}

\section{Introduction} We consider 
 the following  idealized experiment. 
 An amount of a reactant gas of species $A$
is injected into an initially evacuated 
chemical reactor. The reactor, which has arbitrary but well defined shape represented by
a region $U$ in $3$-dimensional space,  is filled with a chemically  inert solid material that is 
permeable to gas diffusion. 
For example, the reactor may be filled with closely packed small inert particles 
creating 
  a network of  channels in the packing interstitial void  through which gas can diffuse.
 A solid catalytic material promoting 
the irreversible  reaction $A\rightarrow B$, where $A$ and  $B$ are gas species that
can diffuse through $U$, is embedded into the packed bed of the reactor
at specified places.  At any given time, the  mixture of reactant and product gases 
is allowed to escape the reactor  through  part of the  boundary of $U$   designated as the reactor exit,
while the outside of $U$ is kept at near vacuum. 
 The composition of the gas mixture $A+B$ exiting  $U$ is analyzed 
and after $U$ is eventually emptied of all gas the molar fraction of $B$ in the total gas outflow is measured.
We refer to this fraction by $\alpha$
and call it
the reaction {\em conversion}, or {\em conversion probability}.  The main concern of this paper is the theoretical determination of $\alpha$ 
for a variety of systems.    More specifically, our focus in this paper is on  the functional dependence of $\alpha$ on the reaction rate constant
and on the geometric configuration of the system, as will be explained in detail shortly. 
The analytical and numerical results described here are based on the general ideas 
developed in \cite{matt1}, which will be summarized later in the paper.

The above  description is meant to capture
the essential features of 
the experimental  technique known as {\em Temporal Analysis
of Products} (TAP) used for studying primarily heterogeneous catalysis involving
gases and complex solid materials. TAP experiments are  based on technology invented by J. Gleaves \cite{tap1} and
 methodology developed by J. Gleaves and  G. Yablonsky \cite{tap3}.  See \cite{wiki} and the references \cite{tap1,tap3,tap4,tap5,tap2}
for a detailed explanation of the TAP method.

In an illustrative  TAP experiment,
one considers $\text{CO}$ oxidation in the presence of   $\text{Pt}$. 
The actual TAP   reactor has the shape of a small cylinder in which  reactant gas is
injected at the center of the circular backside  and product gases exit at the opposite   (open) side.
 The  reactor's inert region is uniformly packed with non-porous small particles of quartz,
creating a medium that can be characterized by essentially constant diffusivity. Diffusion
is generally assumed to proceed in the Knudsen regime, in which 
the mean free path of the gas molecules is comparable or longer than the 
length scale of the small network of voids produced by the packing. 
 It is   natural to model 
the motion of gas molecules  by   mathematical Brownian motion (or Wiener process) as this motion
 corresponds macroscopically to ordinary Fickian diffusion. 
The catalytic zones  may consist of particles having similar size distribution  as 
in the inert region. These zones are kept at
a constant temperature. Although the precise  reaction mechanism resulting in the overall reaction $2\text{CO}+\text{O}_2\rightarrow 2\text{CO}_2$ may be complex, 
one is often justified in representing the kinetic dependence as a simple first order kinetic expression.  In this example,
 the quantity $\alpha$ of interest to us is the molar fraction of $\text{CO}_2$   in
the total gas outflow.

Below is a  summary description of the structure of the paper.
 \begin{itemize}
 \item In section \ref{alphaprocedure} we review an  effective procedure developed in
 \cite{muller,cloninger} and elaborated in  \cite{matt1}  for obtaining conversion $\alpha$ under very
 general conditions, based on solving a time-independent boundary value problem for either Laplace's equation
 in certain cases or, more generally, a Feynman-Kac
 operator.   This procedure was introduced in \cite{cloninger} and  \cite{muller}, and has been developed 
 in much greater mathematical detail
in \cite{matt1}. It is a rather straightforward method that   should be contrasted with the approach of  \cite{tap3},
for example, which depends on first solving 
  the reaction diffusion equations for the concentration of the gas product.  
 \item In many cases of interest the reaction-diffusion system in dimension $3$ can be approximated by one-dimensional
 graph models, in terms    of so-called {\em metric graphs}. For these models
  it is possible to obtain exact  analytic solutions for $\alpha$ as will be seen in section \ref{networks}.
  The basic idea is from \cite{matt1}.  
 For a  graph model having arbitrary topology but
  a single active node,   $\alpha$ depends on the reaction constant\footnote{More precisely, $\kappa$ is  proportional to the reaction constant $k$ in the  way   explained  in Section \ref{alphaprocedure}. } $\kappa$ according to $\alpha=C\tau\kappa/(1+\tau\kappa)$, where
 the coefficients  $C$ and $\tau$ are  independent of   $\kappa$ and depend  on purely geometric features of the graph  including the
  position of the active node, the location of the exit set of nodes, and in the case of $C$ the choice of node of
  gas injection.
  
  \item 
 In section \ref{TAP} we 
 investigate via numerical experiments  some $3$-dimensional examples of
 systems that are similar to TAP-configurations; that is,  cylindrical reactors with one or more catalyst
 pellets.  Our main interest is to determine, in the case of a single pellet placed at varying  positions and
 having a variety of simple shapes,
how well   $\alpha$   can be approximated by the general formula given above for a graph model with
a single active node; and to find out how $\tau$ and $C$ may depend on  
 position and shape parameters of the catalyst pellet. 
This section contains some of the main new results of the paper.
 
  \item  An interesting general problem is to determine the optimal arrangement of catalyst particles
that maximizes $\alpha$. We touch on this optimization problem very briefly by describing what happens
for the linear chain of thin zones example of Figure \ref{thinzones}. We propose a conjectural picture of
how the optimal configuration should in general depend on the reaction constant.

 \end{itemize}

\section{A general procedure for obtaining $\alpha$}\label{alphaprocedure}
This section summarizes the main mathematical facts concerning  the class of   boundary value problems from which the reaction conversion $\alpha$ can be 
obtained. 
More details and the justification of the method can be found in \cite{matt1,cloninger,muller}.
Reference \cite{matt1}, in particular, provides a derivation  based on a stochastic formulation and analysis
of the problem.

 \begin{figure}[htbp]
\centering{}\includegraphics[width=3in]{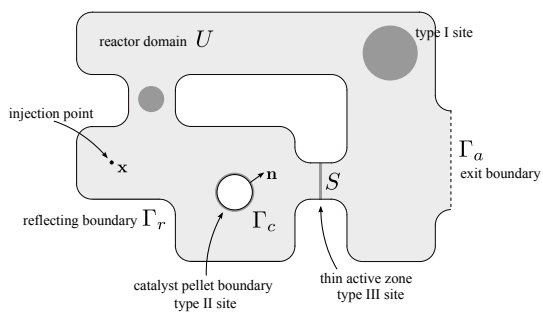} 
\caption{{\small \label{domain} This diagram summarizes some of the notation we use in the boundary value problem for the
determination of the survival function $\psi$. }}
\label{domainschematic}
\end{figure}

We refer to Figure \ref{domain} for a few definitions and notations concerning the domain of diffusion, also referred
to as the {\em reactor}, denoted by $U$. It is convenient to formulate the problem in terms of  the {\em survival function}
  $\psi(\mathbf{x})$,  representing  the probability that a single
gas molecule of type $A$, injected into $U$ at the initial position
$\mathbf{x}$, will eventually leave $U$ through the exit boundary $\Gamma_{a}$ without
having converted into $B$.  
The  complementary probability  is then 
$\alpha(\mathbf{x})=1-\psi(\mathbf{x})$.  We assume that gas molecules diffuse through the reactor bed $U$
under ordinary  (Fickian) diffusion having constant diffusivity, denoted $\mathscr{D}$. 
This amounts to the assumption that the trajectories of individual gas molecules correspond
to sample paths of mathematical Brownian motion.

In real TAP experiments    active particles may consist of a polycrystalline material  forming 
a complex structure of pores and other features at various length scales.  The reaction takes place  on the surface. 
By expressing  $\alpha$ in terms of  solutions to   a boundary value problem as we do below, we
are effectively assuming that all  this complexity  can be reduced to   three basic configurations  of the catalyst material that we refer to as  active sites of types I, II, III:

\begin{enumerate}
\item {\em Active sites of type I.} The active region comprises a subset of  $U$ that is permeable to gas diffusion, where the reactant gas $A$
has constant diffusivity $\mathscr{D}$ equal to that of the inert region of $U$.   Let $q=q(\mathbf{y})$ be the indicator function
of this region,  defined by having  value $1$ if $\mathbf{y}$ lies in the   region and $0$ if not.
The reaction rate is then $k q$ where $k$ is the reaction constant, which has the ordinary physical dimension $1/[\text{time}]$.  

\item {\em Active sites of type II.} The active region is the union of the surface boundaries of one or more solid, non-permeable catalyst particles assumed to have  definite regular shapes.
The interior of each  such  particle  is not regarded as being part of $U$ while the particle boundary surface is supposed to
have a well defined unit normal vector field. We call these  surfaces active sites of {\em type} II.
The reaction constant in this case will be
denoted by $\kappa$, a quantity having  physical dimension $1/[\text{distance}]$. The relationship  between $\kappa$ and $k$ will be discussed below. The former is similar to
a Damk\"ohler number of type II, although we avoid this terminology here and simply refer to $\kappa$ as the {\em surface
reaction constant}. 
\item {\em Active sites of type III.} The active region consists of regular surfaces (having a well-defined normal vector field) that are porous to 
the passage of gas, forming  transition boundaries between inert regions.  We also refer to these sites as {\em active thin zones}.  The 
  reaction  constant for this type of active site will be denoted $\kappa$ and is equal to the constant for sites of
  type II. 
\end{enumerate}

The surface boundary of $U$ is the union of three types of surfaces:
\begin{enumerate}
 \item $\Gamma_a$ is the {\em exit boundary}
through which gas escapes from $U$ (`$a$' stands for ``absorbing''); 
\item $\Gamma_c$ is the union of the active   surfaces of 
type II; 
\item $\Gamma_r$ is the remaining of the boundary of $U$, where
$r$ stands for ``reflecting.''
\end{enumerate}

 \begin{theorem}[See \cite{matt1,muller,cloninger}]
If the overall reaction
constant is $k$ 
 then   $\psi(\mathbf{x})$
satisfies
\[
\mathscr{D}\Delta\psi-kq\psi=0\text{ on }U
\]
 with the following boundary conditions, where $\mathbf{n}$ is a unit normal vector field to $\Gamma_r$: 
\begin{align*}
\mathbf{n}\cdot\nabla\psi & =0\text{ on }\Gamma_{r}\\
\psi & =1\text{ on }\Gamma_{a}.
\end{align*}
The effect of the active regions of types II and III are captured
by a Robin boundary condition on $\Gamma_c$ and by a transition (discontinuity) condition 
for the normal derivative of $\psi$ on active surfaces, respectively.   Thus, for a new constant $\kappa$
to be defined shortly, 
$$
\mathbf{n}\cdot\nabla\psi  =\kappa\psi\text{ on }\Gamma_{c}$$
where $\mathbf{n}$ is the unit normal vector fields on the surface of catalyst particles pointing out into
$U$ and 
$$ \mathbf{n}\cdot \nabla^+\psi - \mathbf{n}\cdot\nabla^-\psi=\kappa \psi$$
on each transition surface $S$, where $\mathbf{n}$ is a normal vector field on $S$ and
$\nabla^\pm \psi$ indicates the limit values of the gradient of $\psi$ when the surface is approached
from behind ($-$) or from the front ($+$), according to the direction of $\mathbf{n}$.
\end{theorem}

It remains to clarify the relationship between $k$ and $\kappa$. The former has physical dimension
of reciprocal of time and the latter reciprocal of distance. Let us 
  imagine that  
 each catalyst particle is surrounded by an active ``collar region,'' or thin shell,   and that 
  each active transition surface $S$ has a definite width,
  so that these active sites can be  regarded as being  of type I. 
  In other words,  there is a zone of catalytic  activity  around each active particle 
  to which we attribute a width $\delta$. 
Similarly for thin zones, or sites of type III.
By  a standard approximation argument using the divergence
theorem, we obtain   the relation
\begin{equation}\label{delta}\kappa=\frac{\delta k}{\mathscr{D}}.\end{equation}

Because the systems we are going to investigate in greater detail
below only involve    active regions of type II and III, it is $\kappa$ rather than $k$
that will appear in our formulas for conversion $\alpha$.  Let us denote
by  $\alpha_\kappa(\mathbf{x})$ the value of conversion for rate constant $\kappa$ and
initial point of gas injection $\mathbf{x}$.  This quantity has the following probabilistic interpretation:
$$\alpha_\kappa(\mathbf{x})=\text{probability that an $A$-molecule initially at $\mathbf{x}$ reacts before leaving $U$}. $$
The limit of  $\alpha_k(\mathbf{x})$ for large values of $\kappa$, denoted
  $\alpha_\infty(\mathbf{x})$,  is the probability that the gas molecule started at $\mathbf{x}$
visits an active site before leaving the reactor. This hitting probability is a well-studied quantity in 
probability theory that can be obtained by solving a Dirichlet boundary value problem for Laplace's equation.
Thus our main concern here is the determination of   $\alpha_\kappa(\mathbf{x})/\alpha_{\infty}(\mathbf{x})$.
This quantity may be interpreted as the {\em conditional conversion}, or the conversion probability of a reactant molecule conditional on the
molecule actually hitting the catalyst.

\section{Metric  graph  approximation} \label{networks} Although the aforementioned procedure for computing $\alpha$
can be implemented numerically quite  effectively (examples will be given below), it is  useful
to have model configurations of the reaction-diffusion system that can be solved analytically.
We are particularly interested in determining explicitly how $\alpha_\kappa(\mathbf{x})/\alpha_\infty(\mathbf{x})$
depends on  $\kappa$.  Such explicit formulas can then  serve as
a basis of comparison to guide the investigation of more complicated systems.

     \vspace{0.1in}
\begin{figure}[htbp]
\centering{}\includegraphics[width=2.5in]{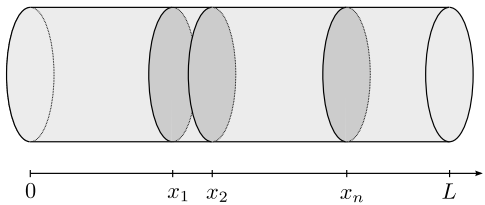} 
\caption{{\small A cylindrical reactor with $n$ active thin zones of type III. Gas is injected at position
$0$ along the axis of the cylinder and the exit is the open side at $L$. We call this the {\em linear chain of thin zones} example with
$n$ catalyst sites.}}
\label{thinzones}
\end{figure}
    
 In this section we explore a class of model systems defined on metric graphs.   (See
  \cite{matt1} for more details.) Let us first consider the relatively simple example indicated in Figure
  \ref{thinzones} to illustrate the main ideas.

 \vspace{0.1in}
\begin{figure}[htbp]
\centering{}\includegraphics[width=2.5in]{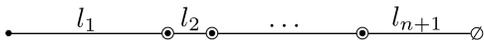} 
\caption{{\small The metric graph representing the example of Figure \ref{thinzones}. Here $l_j=x_j-x_{j-1}$.  Edges
  are labeled with lengths, $\emptyset$  indicates an exit node, a simple dot indicates an 
inert  node, and $\odot$ represents an active (permeable) node. Nodes connected to 
a single edge,  like the leftmost one in this figure, are reflecting, except for exit nodes.} }
\label{graphthinzones}
\end{figure}

If we make the simplifying assumption that gas is injected   over the part of the reactor  boundary
at   coordinate $x=0$ along the cylinder axis (the circular backside; see Figure \ref{thinzones}), uniformly over the points
of that backside, then by symmetry the solution of the boundary value problem for conversion 
only depends on the variable  $x$. The problem thus reduces to
dimension $1$.  See Figure \ref{graphthinzones}.

 \vspace{0.1in}
   \begin{figure}[htbp]
\begin{centering}
\includegraphics[width=2.0in]{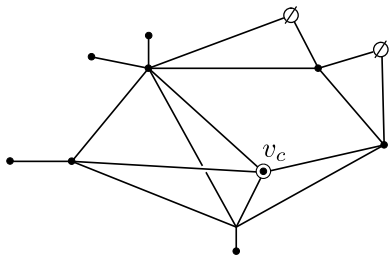}
\caption{{\small{}\label{singlecat} A network model with one catalyst particle at the node $v_c$. See  Figure \ref{thinzones}}
for the convention we are using to represent the different types of nodes.}
\end{centering}
\end{figure}

More generally, we may consider network  models as in Figure \ref{singlecat} specified by the following data: a set $\mathscr{V}$ of nodes (or vertices)
which contains a subset $\mathscr{V}_c$ of active nodes and the set $\mathscr{V}_\emptyset$ of
exit nodes, all other nodes being inert; and a set $\mathscr{E}$ of edges connecting pairs of nodes, each edge assigned
a length. In our network diagrams inert vertices are indicated by a dot ``$\cdot$'', active nodes by $``\odot$'',
and exit nodes by ``$\emptyset$''. 
   If $e$ is an edge, we let $|e|$ denote its length. It is convenient to introduce the function $q(v)$ on the
set of nodes such that $q(v)=1$ if $v\in \mathscr{V}_c$ is an active node, and $q(v)=0$ if $v$ is inert.  For each
node $v$ let $\mathscr{E}_v$ indicate the set of edges having $v$ as one of its two endnodes. If $e$ is an element
of $\mathscr{E}_v$ then $r(e)$ will be used to indicate the other end of $e$ opposite to $v$. The {\em degree} $\text{deg}(v)$
of a node $v$ is defined as the number of edges issuing from $v$. 
With this notation in place, the boundary value problem for the survival function $\psi$ restricted to $\mathscr{V}$
reduces to the system of linear equations:
\begin{equation}\label{general}
\sum_{e\in \mathscr{E}_v}|e|^{-1} \psi(r(e))= \left(\kappa\,  \text{deg}(v) q(v) + \sum_{e\in \mathscr{E}_v}|e|^{-1}\right)\psi(v) \text{ if $v$ is not in }   \mathscr{V}_\emptyset
\end{equation}
and 
\begin{equation}
 \psi(v)=1 \text{ if $v$ lies in } \mathscr{V}_\emptyset.
 \end{equation}
 For more details, see \cite{matt1}. Thus, for such network models,  finding $\alpha_\kappa(\mathbf{x})=1-\psi(\mathbf{x})$,
 where $\mathbf{x}$ is now a node of the metric graph, reduces to the elementary problem of solving a system of linear equations. 
 
 It is shown in \cite{matt1} that the solution to this system can be expressed as follows. For each $v\in \mathscr{V}_c$
 there is a polynomial $\tau_v(\kappa)$ in  $\kappa$ of degree less than the number of active nodes such that
 $$ \alpha_\kappa(\mathbf{x})= P(\mathbf{x}) -\sum_{v\in \mathscr{V}_c} P_v(\mathbf{x})\frac{\tau_v(\kappa)}{\tau(\kappa)}$$
 where $\tau(\kappa)$ is a polynomial in $\kappa$ of degree less than or equal to the number of active nodes, $P(\mathbf{x})$
 is the probability that a diffusing molecule  started at $\mathbf{x}$ hits the set $\mathscr{V}_c$ before reaching
 the exit set $\mathscr{V}_\emptyset$, and $P_v(\mathbf{x})$ is the probability that the molecule hits the active set
 first at $v$. We give in the next subsections  a few concrete examples.

  Metric graphs, also known as quantum graphs,  have been extensively used to describe a variety of phenomena. A  detailed  survey of
 the literature   can be found in \cite{kuchment}.  We emphasize that, in the present application, diffusion is restricted to
  the edges of the graph, hence it is fundamentally $1$-dimensional. This is the essential simplification afforded by these systems. 
  On the other hand, more complicated higher dimensional systems can be approximated by such $1$-dimensional systems if
  the graph architecture is suitably chosen. 
  
 \subsection{General solution for a single active site}
 It is shown in \cite{matt1} that for the special case of  network models with a single catalyst node 
 the following general expression for $\alpha_\kappa(\mathbf{x})$ holds:
 \begin{equation}\label{single}\alpha_\kappa(\mathbf{x})=P(\mathbf{x},v_c)\frac{\tau\kappa}{1+\tau\kappa}\end{equation}
 where $v_c$ is the node containing the catalyst and $\mathbf{x}$ is any point of gas injection. See Figure \ref{singlecat}.
 The quantity $P(\mathbf{x},v_c)$ is the probability that the diffusing molecule started at $\mathbf{x}$ will hit $v_c$ before
 leaving at one of the exit nodes indicated in the figure by $\emptyset$.  
 
\begin{figure}[htbp]
\begin{centering}
\includegraphics[width=5.2in]{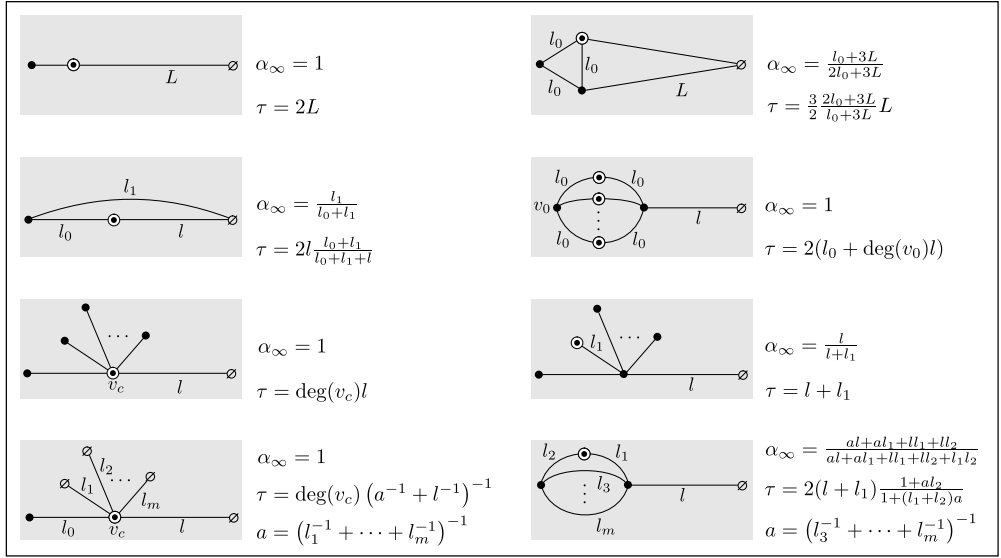}
\caption{{\small{}\label{bestiary} A short bestiary of elementary network models. Recall: $\alpha_\infty$ is the probability that
a diffusing molecule hits a catalyst node; $\tau$ is the parameter in $\alpha_\kappa/\alpha_\infty=\tau\kappa/(1+\tau\kappa).$
The label near an edge is its length; $\odot$ indicates a catalyst node, $\emptyset$ is an exit node
and $\cdot$ is  an inert node, reflecting if its degree is $1$.  The point of injection is the leftmost node. These examples
are all described by  Equation \ref{single}.}}
\end{centering}
\end{figure}
 
 The quantity $\tau$ is purely geometric and does not depend on $\kappa$. It contains information
 about the lengths of edges, degrees of nodes and the positions of $v_c$ and the exit nodes. We recall that the {\em degree} of a node
 $v$, denoted $\text{deg}(v)$, is the number of edges issuing from that node.
 Roughly speaking, $\tau$ may be viewed as
 the expected total   time   that the diffusing molecule spends at $v_c$ conditional on the molecule actually
 passing through $v_c$.\footnote{Clearly, $\tau$ cannot be an actual  time parameter as it has physical dimension of distance. The precise meaning  of $\tau$ requires the probabilistic notion of {\em local time}
 and is explained in \cite{matt1}.}
We may write \ref{single} alternatively as 
\begin{equation}\label{basic}\frac{\alpha_\kappa(\mathbf{x})}{\alpha_\infty(\mathbf{x})}=\frac{\tau\kappa}{1+\tau \kappa}.
\end{equation}

Some  examples of model networks    are shown 
in Figure \ref{bestiary}. Their respective $\alpha_\kappa$  are obtained by solving the linear system \ref{general}
 and are shown to satisfy   Equation \ref{basic} with the  indicated  values of $\alpha_\infty$ and $\tau$.
 The point of injection $\mathbf{x}$ is the leftmost node in all cases.

For example, for  the third graph from the top on the left column of Figure \ref{bestiary}
$$\alpha_\kappa= \frac{\text{deg}(v_c)l\kappa}{1+\text{deg}(v_c)l\kappa}. $$
Note  that conversion, in this example,  grows with the number 
$\text{deg}(v_c)$
of edges issuing from the active node, although only the length of the edge leading to the exit
node enters into the formula. The other edges have the same effect no matter how long or short they are, 
so long as their length is positive.  Said differently:

{\em By  adding reflecting, inert  nodes in the vicinity of an active  node we   increase the residence time at the latter node and consequently increase conversion for the  network as a whole. 
}
 
  \begin{figure}[htbp]
\begin{centering}
\includegraphics[width=3.5in]{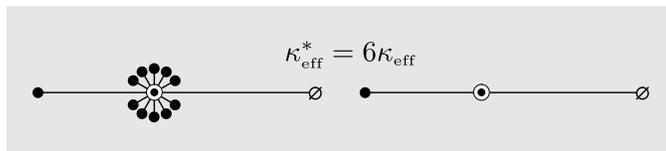}
\caption{{\small\label{star} In both  graphs, the single active node has the same reaction constant $\kappa$ but the effective constant, which we define as $\kappa_{\text{\tiny eff}}=\tau\kappa$, is
six times greater for the graph on the left, whose constant is denoted $\kappa_{\text{\tiny eff}}^*$. More generally, $\kappa^*_{\text{\tiny eff}}=\frac12 \text{deg}(v_c) \kappa_{\text{\tiny eff}}$, where $\text{deg}(v_c)$ is the degree of the active node of the graph on the left.  }}
\end{centering}
\end{figure}

 This property may suggest a way to design the active units so as to increase their activity. See  Figure \ref{star}. 
 In this  example, the active unit consists of one active node surrounded by inactive ones forming a system of ``spokes.''
 Note that diffusion can only occur along the spokes connecting   the active node to each of the inactive nodes. The
 effect  is to greatly increase the residence time (technically, the local
 time of the Brownian path at the active node) thus increasing the effective level of activity of the active unit. It is interesting to observe that
 the length of the spokes is irrelevant since only their number, or the degree $\text{deg}(v_c)$ of the active node, enters into the expression of conversion
 for this example.

 \subsection{Conversion for the linear chain of thin zones model}
 The reader is referred to the linear chain of thin zones example of  Figure \ref{thinzones} and its
  network model of Figure \ref{graphthinzones}. We assume that the point of gas injection is the leftmost
  node (at $x=0$). Let $n$ be the number of catalyst nodes and $l_1, \dots, l_{n+1}$ the edge lengths,
  so that $l_j=x_j-x_{j-1}$. The system of equations for the survival function $\psi(v_j)$ evaluated at
  the nodes $v_0, v_1, \dots, v_{n}$ are in this case
  $\psi(v_0)=\psi(v_1), \psi(v_{n+1})=1$ and $$ \frac{\psi(v_{j+1})-\psi(v_j)}{l_{j+1}}=  \frac{\psi(v_{j})-\psi(v_{j-1})}{l_{j}} + k\psi(v_j) \text{ for } j=1, \dots, n.$$ 
  Important:  in this subsection only, we have chosen to think of $\delta$ (see the relation \ref{delta}) as the width of an active region
  around a node rather than the radius in order to conform to  the discussion in \cite{cloninger} and elsewhere. Therefore, $\kappa$ here
  is actually half of  the $\kappa$ appearing in Equation \ref{general}.
 For other networks  we let our $\kappa$ be again as in Equation \ref{general}, which we think is a more natural convention.

 It can  then be shown, by solving the above system, that $$\alpha_\kappa(0)= \frac{f_{n+1}-1}{f_{n+1}}$$
  where  $f_{n+1}$ is the last number in the sequence $f_1, f_2, \dots, f_{n+1}$ obtained recursively by
  $$f_1=1,\ \  f_{j+1}= f_j+ l_{j+1} \left(f_1+\cdots + f_j\right) \kappa, \text{ for } j=2, \dots, n.$$
  Thus, for a single catalyst site, the conversion probability is
 $$\alpha_\kappa(0)= \frac{l \kappa}{1+l\kappa} $$
 where $l$ is the distance from the  catalyst node to the exit at $x=L$. 
 For two catalyst nodes we obtain
\begin{equation}\label{twonodeformula}\alpha_\kappa(0)= \frac{(l_2+2l_3)\kappa + l_2l_3\kappa^2}{1+(l_2+2l_3)\kappa + l_2l_3\kappa^2}. \end{equation}

  We may ask in this case: What is the optimal configuration of nodes? That is, what should be the
  values of $l_j$ that maximize conversion? Note that $\alpha_\kappa(0)$ increases monotonically with
  $f_{n+1}(\kappa)$ as a function of $\kappa$. When $\kappa$ is small relative to the lengths $l_j$ then
  it is easily seen that $f_{n+1}(\kappa)= 1 + (l_2+2l_3+\cdots + n l_{n+1})\kappa + \text{higher order terms in $\kappa$}. $
  This expression is equivalent to
  $$f_{n+1}(\kappa)= 1 + \left[(L-x_1)+ (l-x_2)+ \cdots + (l-x_n)\right]\kappa + \text{higher order terms in $\kappa$}. $$

  On the other hand, when $\kappa$ is relatively large,
  $$ f_{n+1}(\kappa)= l_2l_3\cdots l_{n+1} \kappa^n+\text{lower order terms in $\kappa$}.$$ 
  The configuration that maximizes the 
  coefficient 
  of $\kappa^n$   corresponds to $x_1=0$ and
  $$ x_{j}= \frac{x_{j-1}+x_{j+1}}{2} \text{ for } j=2, \dots, n.$$

  Thus we are led to the following qualitative  conclusion:
  
   {\em  When $\kappa$ is sufficiently small,
  conversion is maximized by having all the
  catalyst nodes   bunched together right at the point of gas injection.  When $\kappa$ is very large, the optimal configuration
  for conversion approaches that in which the catalyst nodes are equally spaced along the line of the reactor.}

\begin{figure}[htbp]
\begin{centering}
\includegraphics[width=3.5in]{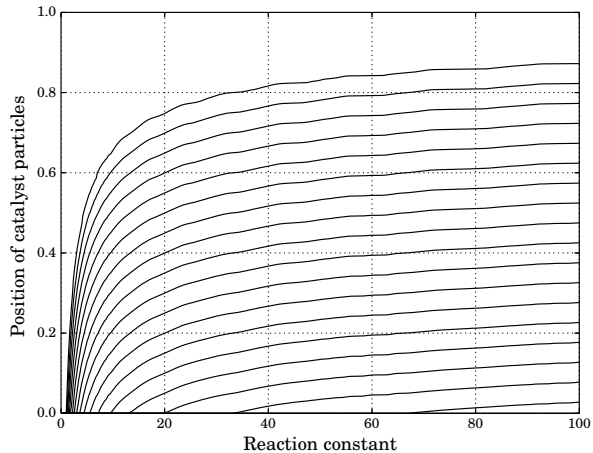}
\caption{{\small{}\label{optimal}Optimal configuration of a multi-particle thin zone system.  For each value of $\kappa$ along
the horizontal axis   the vertical axis gives the position of each catalyst particle. Note how the particles
one by one detach  from the bunch at position $0$ as $\kappa$ increases. For any given value of $\kappa$, those
particles that have already detached lie at fairly uniform distance from each other. }}
\end{centering}
\end{figure}

This phenomenon is illustrated in Figure \ref{optimal} for a linear chain system with $20$ catalyst sites.
It is interesting to observe how, for each given value of $\kappa$, the catalyst nodes that have
already detached themselves from the group bunched at $0$ are fairly equally spaced among themselves. 

   \vspace{0.1in}

\section{Numerical experiments} \label{TAP}
In this section we obtain numerically the conversion $\alpha$ for a variety of $3$-dimensional reactor configurations 
that can potentially be studied  experimentally in so-called TAP (Temporal Analysis of Products)-experiments. 
The   objective here is two-fold: to begin a detailed examination of how $\alpha$ depends on geometric parameters
of the system such as position, shape, and distribution of the catalyst particles; and the extent to which
formulas derived for  network models can be used to approximate the dependence of  $\alpha$ on 
the rate constant $\kappa$,  for a class of  $3$-dimensional
systems. 

More specifically, we have seen that for network models with a single active node the relation
$\alpha_\kappa/\alpha_\infty=\tau\kappa/(1+\tau\kappa)$ always holds, where $\tau$ is a constant
independent of $\kappa$ characteristic of the geometric configuration of the system.\footnote{As explained earlier, $\tau$ 
may be regarded as a proxi for  the molecular expected residence time at the active node of the graph,
conditional on the molecule passing by that node. The precise probabilistic  interpretation of $\tau$ as
a {\em local time} is given in \cite{matt1}.} Thus we 
wish to investigate whether the same dependence on $\kappa$ holds for $3$-dimensional configurations with
a single, relatively small, catalyst particle.  It will be seen that this functional dependence on $\kappa$
does hold very satisfactorily. On the other hand, the relationship between network models with more than
one active node and $3$-dimensional systems with more than one catalyst particle is far from
obvious.

\begin{figure}[htbp]
\begin{centering}
\includegraphics[width=5in]{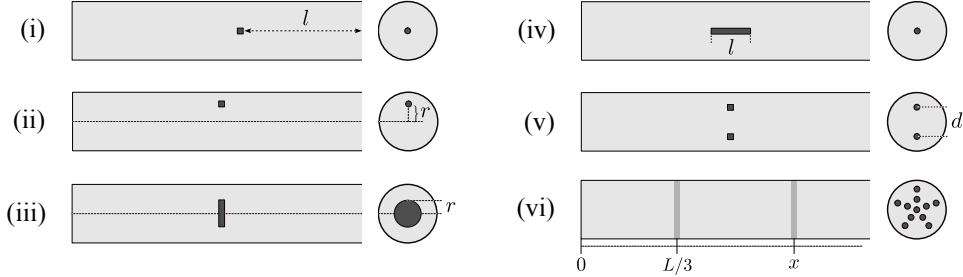}
\caption{{\small{}\label{six}    Six reactor configurations in dimension $3$ that we  investigate numerically in this section. The point
of injection in all cases is the middle point of the back (left) side of the cylinder.}}
\end{centering}
\end{figure}

Figure \ref{six} shows the six configurations we wish to investigate.  In every case, the reactor is a cylinder,  with the dimensions exactly as shown
in the figure, having ratio of length over diameter equal to $5$. There may be  one or more catalyst particles, which also have  a cylindrical shape of varying lengths, radii, positions,  and number, and
the particles    correspond to active regions of type II as defined in Figure \ref{domain}.

 The experiments differ as follows:
  Experiments (i), (ii), (iii), and (iv) are single particle systems; here the main goal is to test the validity of  formula \ref{basic}
  as far as the dependence on the reaction constant is concerned. In experiments (iii) and (iv), the size of the particle changes
  and the focus is on
  how large it can be before the formula for single active node network  conversion is no longer valid.
  Experiment (v) is  a two-particle system that is still fairly well approximated by the single node graph model, as will be seen. Finally, experiment (vi)
  involves two groups of particles. The goal in this case is to compare with the linear chain network with two active nodes.
  
 In experiments (i), (ii), (iii),
 (v), and (vi) the catalyst particles
 all have ratio of length over diameter equal to $1$, and their
diameter is $10$ times smaller than the diameter of the reactor.  Only in experiments (iii) and (iv) are the shapes of the
particles changed.
 
The     boundary value problems in each of the experiments was  solved using  the computational software  FEniCS running
on an ordinary PC. In each experiment we give plots for $\alpha_k, \alpha_\infty$ and for $\tau$ as a function of the changing parameter
of the experiment; for example, the distance from the particle to the exit in experiment (i). The criterion to test the validity of
the network model formula  for a single active node is based on the relative error in $\tau$ (see graphs in Figure \ref{errors}).
This is explained   below.

\subsection{Experiment i: changing particle position on reactor axis}
\begin{figure}[htbp]
\begin{centering}
\includegraphics[width=3.9in]{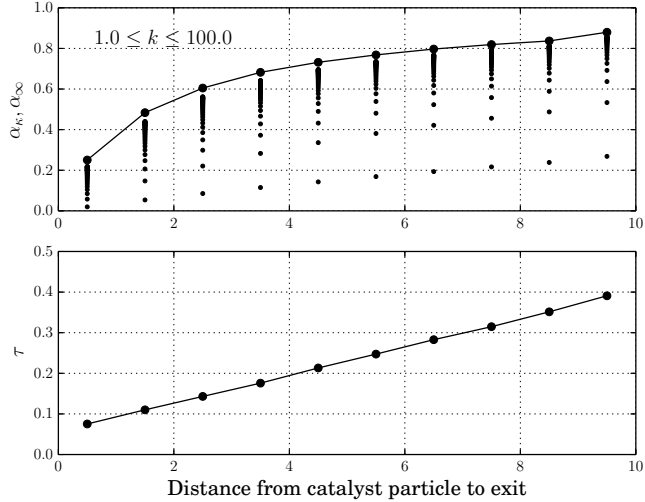}
\caption{{\small{} \label{exp1}Experiment for configuration (i) of Figure \ref{six}. The
parameter of interest is the distance  of the catalyst particle from
the reactor exit. }}
\end{centering}
\end{figure}

\begin{figure}[htbp]
\begin{centering}
\includegraphics[width=4.5in]{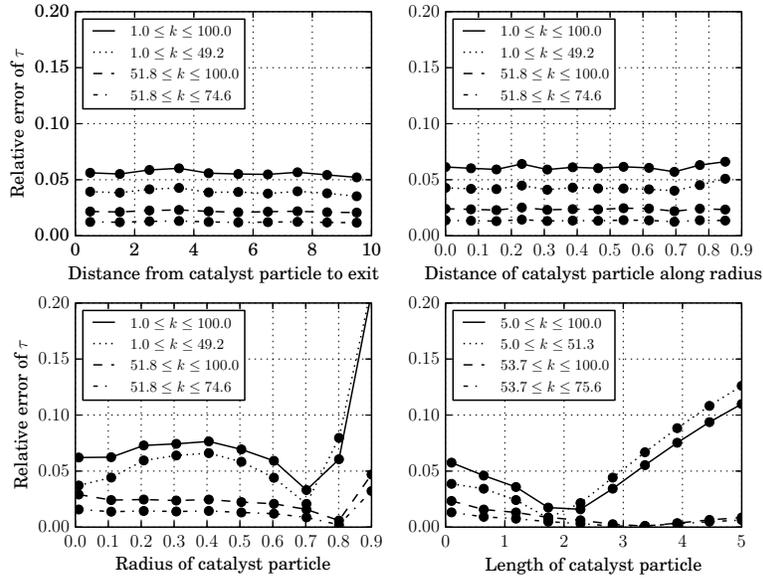}
\caption{{\small{}\label{errors} The relative error of $\tau$, as defined in the text, measures the degree to which
$\tau_{\text{\tiny exp}}$ is independent of $\kappa$. A small value for this error indicates that
the single particle network model is a good approximation for the $3$-dimensional systems. }}
\end{centering}
\end{figure}

Consider the configuration (i) shown in Figure \ref{six}. A single catalyst particle is
placed along the central axis of the reactor. The parameter of interest is the distance $l$ from the particle
to  the reactor  exit   on the right end.  In this case we wish to know how well 
the expression $\alpha_\kappa/\alpha_\infty=\tau\kappa/(1+\tau\kappa)$ holds and whether
$\tau$ depends linearly on $l$ as in   the    linear chain of active nodes of Figure \ref{graphthinzones} but  with a single node. 
 The   graphs of   Figure \ref{exp1}
 and the top-left graph of Figure \ref{errors} show the main results.

 The explanation we give here for how to interpret those graphs  
 will also  apply  to the corresponding  graphs of  experiments (ii), (iii), (iv), and (v). 
 First consider   the top graph of Figure \ref{exp1}. It  gives values of $\alpha_\kappa$
(the vertical line of dots above each value of $l$ on the $x$-axis) for the range of $\kappa$ indicated near  the top left corner. 
We  tested  $10$ values of $l$ in the interval from $0$ to $10$ (the length of the
reactor is $10$ in arbitrary units);  and
$40$ values of  $k$  evenly spaced from  $1.0$ to  $100.0$ for each value of $l$. 
Note that small values of $l$  correspond to catalyst positions
close to the exit side of the cylindrical reactor.  
The
graph drawn in solid line on the same plot (towards which the dots accumulate) gives $\alpha_\infty$ as a function of $l$. Recall that $\alpha_\infty$ is the probability that a gas molecule actually
hits the catalyst, whether or not a reaction occurs. This is the same as conversion when the reaction constant is infinite.

 As expected, conversion increases as the catalyst particle
is placed closer to the injection point. More notable is the lower graph of Figure \ref{exp1} showing
the value of $\tau$ as a function of $l$. The plotted value of $\tau$ (that we call, {\em experimental} $\tau$)   is {\em defined} 
 by
\begin{equation}\label{tauexp}\tau_{\text{\tiny exp}}= \frac1\kappa\frac{\frac{\alpha_\kappa}{\alpha_\infty}}{1-\frac{\alpha_\kappa}{\alpha_\infty}}. \end{equation}
In words, we define $\tau_{\text{\tiny exp}}$   as the value   the data would give for $\tau$ under the assumption that
the single active node graph model were valid. Because for single active node graphs $\tau$ is independent of $\kappa$,
the degree to which $\tau_{\text{\tiny exp}}$ is insensitive to changing $\kappa$ is a measure of the validity of
formula \ref{basic} in  the $3$-dimensional situation.\footnote{For simplicity, we write $\tau$ rather than $\tau_{\text{\tiny exp}}$ on
the graphs.}

To determine whether the experimentally obtained $\tau$ is approximately constant in $\kappa$ we look at 
the top-left plot of Figure \ref{errors}. The quantity plotted as a function of $l$, which we
call the {\em relative error of $\tau$}, is
defined as the ratio of the standard deviation of $\tau$ over the maximum value of
$\tau$ for the indicated ranges of $\kappa$. For example, the top graph (in solid line) of
the top-left plot in Figure \ref{errors} is obtained as follows: for each of the $10$ values of $l$
we compute the standard deviation of the values of $\tau$ corresponding to $1.0\leq \kappa\leq 100$
(the full range of $\kappa$) and divide this standard deviation by the maximum value of $\tau$ over this range.

We do the same for a few restricted ranges of $\kappa$  for an indication of
where the network approximation is better or worse.  For example, 
the  network approximation holds best for the middle range of $\kappa$,
between approximately $50$ and $70$, although it  seems quite reasonable to conclude that
the network approximation is   justified by this single particle experiment over most values of $\kappa$ less than
$100$.

\subsection{Experiment ii: varying the radial distance}
Here we wish to determine to what extent $\tau$ is affected by varying the
position of the catalyst particle along the reactor's  radial direction. The particle is assumed 
to lie at the middle  point  along the central axis coordinate ($l=5$) while
 the radial coordinate of the particle's center, denoted $r$, varies from $0$ to $0.85$. 
 Note that the last value of $r$ gives a gap between the particle and reactor wall  
 of $0.05$. 
   See the second diagram of Figure \ref{six}.

 The first conclusion drawn from examining  the top-right graph of Figure \ref{errors}
 is that Equation \ref{basic} for the dependence of $\alpha$ on $\kappa$ seems to hold as
 well here as it did in the first experiment.  Recall that the relative error measures
 the failure of $\tau$ to be independent of $\kappa$; small values of this error
 supports the validity of approximating the $3$-dimensional system by a single active node network model as
 far as the dependence on $\kappa$ is concerned.

 \begin{figure}[htbp]
\begin{centering}
\includegraphics[width=4in]{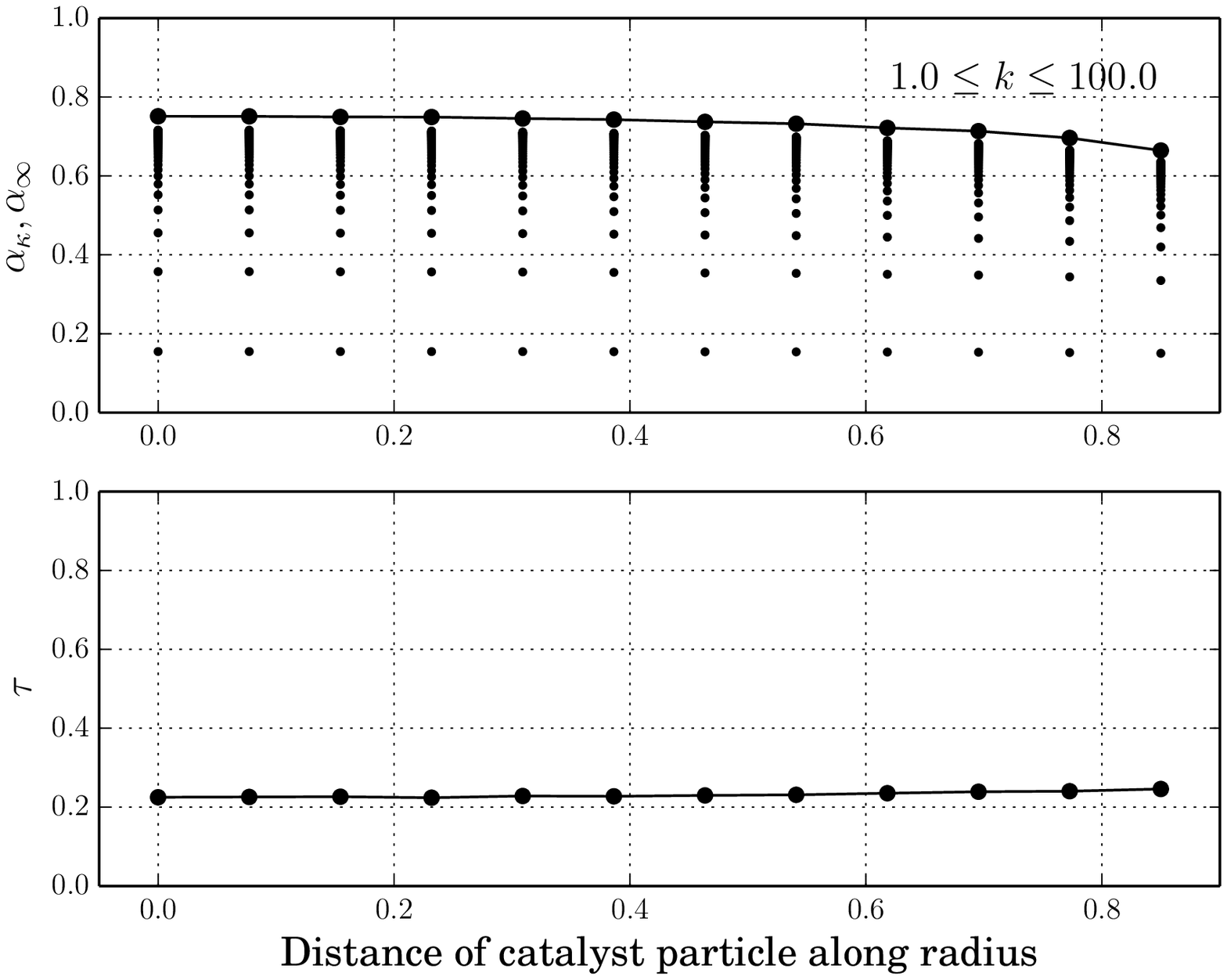}
\caption{\label{experiment2}{\small{} The catalyst particle is placed at the middle of the reactor at a distance $r$ from
the central axis. We computed $\alpha$ for $12$ equally spaced values of $r$,  and for each of these  values 
$\alpha$ was obtained for the same range and number of values of $\kappa$ as in the first experiment. We observe that
$\tau$ is nearly insensitive to $r$, although the hitting probability $\alpha_\infty$ decreases slightly
as $r$ increases. }}
\end{centering}
\end{figure}

  The dependence of $\alpha$ on $r$ is shown on
 the top graph of Figure \ref{experiment2}. Note that conversion decreases slightly
 as the particle is placed closer to the wall of the reactor. At the same time, $\tau$
 does not appear to be much affected by the change in $r$. This seems
 to indicate that varying $r$ mainly affects the
 probability that gas molecules hit the catalyst.

\subsection{Experiment iii: varying the radius of the particle}
In this experiment the position of the catalyst particle is fixed at the middle point
along the central axis of the reactor while   the parameter of interest is the particle's diameter.
We sample $10$ equally spaced values of this parameter,  from $0.01$ to $0.9$. The same $40$ values of $\kappa$ as in the previous two experiments are
also used here.

The bottom-left graph in Figure \ref{errors} gives the relative error in
$\tau$ as a function of the ratio of the radius of the particle over the radius of
the reactor. For values of this ratio less than approximately $0.8$  the
error is comparable with those for the previous two experiments, but it greatly increases
when the particle approaches the diameter of the reactor and becomes a significant 
obstacle to the passage of gas molecules. This means, in effect, that 
a   catalyst particle of large diameter  cannot be reasonably modeled by a single catalyst node
network system (a single thin-zone model). It is not clear, however, why   the relative error for values of the radius between 
$0$ and $0.7$ should have the shape shown. Also note that the main contribution to the relative error in $\tau$
comes from the  lower range of values of $\kappa$. Over the range $50<\kappa<100$
the relative error is comparable to that of the previous two experiments.

  \begin{figure}[htbp]
\begin{centering}
\includegraphics[width=4in]{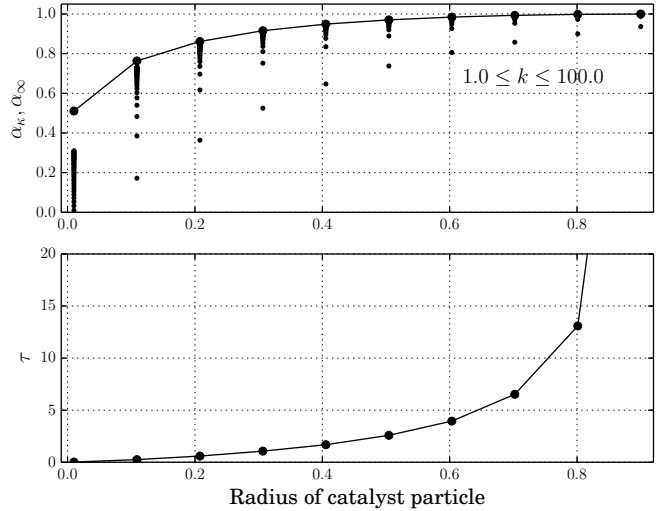}
\caption{\label{experiment3}{\small{}For the highest value of $r$ the
catalyst particle  nearly blocks the cylinder; the value of the corresponding $\tau$ is approximately $50$. }}
\end{centering}
\end{figure}

The dependence of $\alpha_\kappa$ and $\alpha_\infty$ on the catalyst particle radius shown in Figure \ref{experiment3} is,
as should be expected, an increasing function. Also $\tau$ increases significantly with this
radius.  Recall that $\tau$ serves as a proxi for the expected time a gas molecule spends in
the vicinity of the catalyst, assuming that it actually hits the catalyst.  With this in mind,
it is not surprising that $\tau$ should be an increasing function of the radius with no upper bound.  
  
  \subsection{Experiment iv: varying the length of the particle}
\begin{figure}[htbp]
\begin{centering}
\includegraphics[width=4in]{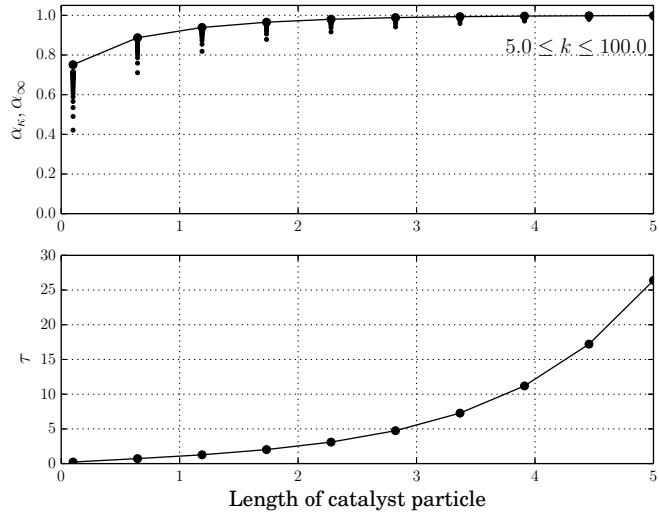}
\caption{\label{experiment4}{\small{} The center of the catalyst particle is fixed at the middle
point of the reactor while its length changes from $0.1$ to $5.0$. The diameter of the
particle remains constant at one-tenth of the diameter of the reactor.  }}
\end{centering}
\end{figure}

  In this experiment we fixed the center of the  catalyst particle at the middle point 
  of the reactor along the center axis and changed the length of the particle from
  $0.1$ to $5.0$ (the latter is half the length of the reactor) in $10$ equally spaced values.
  The diameter of the particle remained fixed at one-tenth of the diameter of the reactor.  
  The graphs of Figure \ref{experiment4} and the lower-right graph of Figure \ref{errors}
are based on $40$  equally spaced values of $\kappa$
between $5.0$ and $100.0$,
 for each value of the length parameter.

The first observation to make is that the thin-zone network model approximation seems
to hold reasonably well for catalyst lengths up to about $3$ (compared to the reactor length $10$)
as can be seen in Figure \ref{errors}.
For greater particle lengths the relative error increases significantly and the single particle network
system  is no longer a good model. Conversion $\alpha$ and the parameter $\tau$, unsurprisingly,  
increase as the length of the particle increases as shown in Figure \ref{experiment4}.

 \subsection{Experiment v: two particles at the center of the reactor}
 We now consider two catalyst particles with central axis coordinate equal to half the
 length of the reactor, placed symmetrically about that axis. We let $d$ denote the distance between the
centers of  two particles. Thus $d$ varies from $0.1$ to $0.9$, where reactor diameter is set to $1$. (Recall
that the diameter of the catalyst particle is  one-tenth of the reactor diameter.)
 
\begin{figure}[htbp]
\begin{centering}
\includegraphics[width=4in]{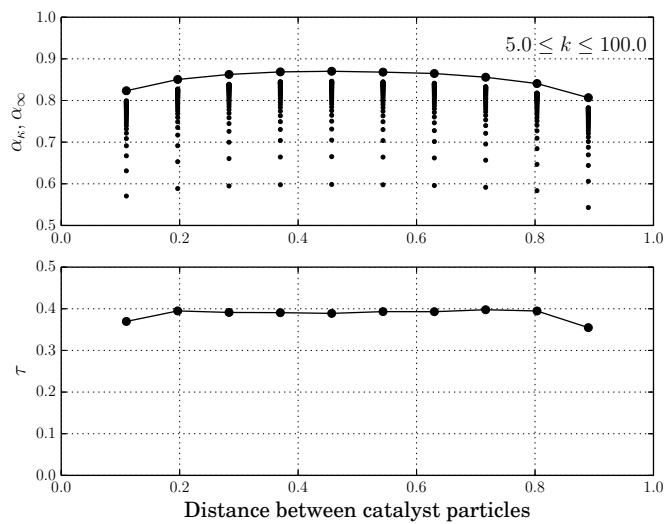}
\caption{\label{experiment5}{\small{}The two catalyst particles lie at the middle point of the reactor, symmetrically
positioned along the radial direction as shown in Figure \ref{six}. Note that $\alpha$ is maximized when
the two particles are at a distance from each other approximately equal to half the reactor diameter.}}
\end{centering}
\end{figure}

\begin{figure}[htbp]
\begin{centering}
\includegraphics[width=2.5in]{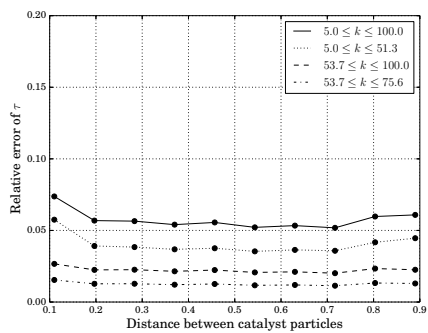}
\caption{\label{experiment5error}{\small{}Relative error for the experiment corresponding to
  configuration (v) of Figure \ref{six}. }}
\end{centering}
\end{figure}

 The graphs of Figure \ref{experiment5} and Figure \ref{experiment5error}  were obtained for
 $10$ equally spaced values of $d$ between $0.1$ and $0.9$ and for each such value 
 we used the same range and number of values of $\kappa$ as in experiment (iv).
The relative error shown in Figure \ref{experiment5error} is comparable to those of experiments (i) and (ii). They are  relatively small, which
  suggests again that the single thin-zone network model captures reasonably well the 
 behavior of this system. 

One noteworthy feature of the upper graph of Figure \ref{experiment5} is that it 
shows the existence of an optimal distance $d$ between the particles at which  conversion $\alpha$ is maximized 
for all values of $\kappa$.  Maximal conversion is obtained when $d$  is approximately half the reactor  diameter. The parameter $\tau$, on the other hand,
is not much affected by varying $d$, suggesting that the change in $\alpha$ is mainly due to
the change in hitting probability, $\alpha_\infty$. 

This behavior may seem somewhat surprising given that, individually, each particle maximizes $\alpha$ when
they are located on the central axis as the upper graph of Figure \ref{experiment2}  indicates. (The effect
shown in that experiment is small but clearly noticeable.)  This suggests that the particles 
  subtract from each other's effect when they are   too close. 
This ``interference''  property is  a bit subtle.  
Recall  from our analysis  of the multi thin-zone network model how the optimal arrangement of catalyst particles
depends on the value of $\kappa$, as shown in Figure \ref{optimal}.  This point merits further investigation. 

\subsection{Experiment vi: two groups of particles}
We turn now to multiparticle configuration (vi) of Figure \ref{six}. There are $2N$ catalyst particles
arranged into two groups of $N$ particles; one group is placed along the central axis a distance from the
injection point equal to one-third of the length of the reactor
so as to form a set of equally spaced spokes. The second group is
similarly arranged at a distance $x$ from the injection point. When $N$ increases, the
system is expected to approximate a two thin-zones system with catalyst zones of type III
according to the classification of zones of Figure  \ref{domain}. 

 \begin{figure}[htbp]
\begin{centering}
\includegraphics[width=3.5in]{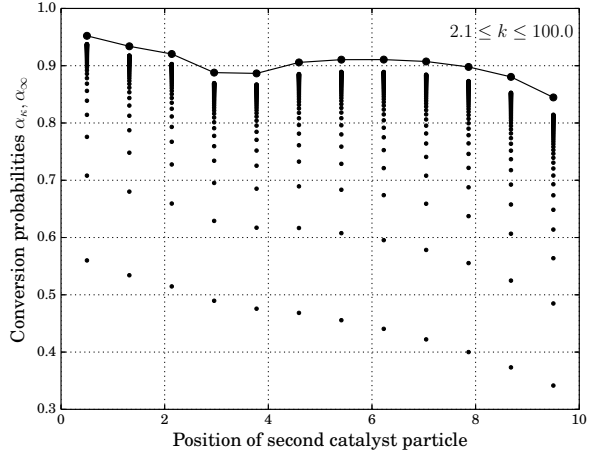}
\caption{\label{twozonealpha}{\small{}
Graph of conversion for the configurations described in Figure \ref{six} (vi) with a single particle in each group.
The first particle is placed at position $10/3$ where $10$ is the length of the reactor in arbitrary units. 
 }}
\end{centering}
\end{figure}

The natural  question here is
whether the linear chain graph model with two active nodes is a good approximation of this $3$-dimensional system
as far as the dependence on $\kappa$ is concerned.  This is equivalent to asking whether the
$3$-dimensional thin zones system of Figure \ref{thinzones}, with two zones, and the $3$-dimensional system with
two groups of particles behave in a similar way.

\begin{figure}[htbp]
\begin{centering}
\includegraphics[width=5.5in]{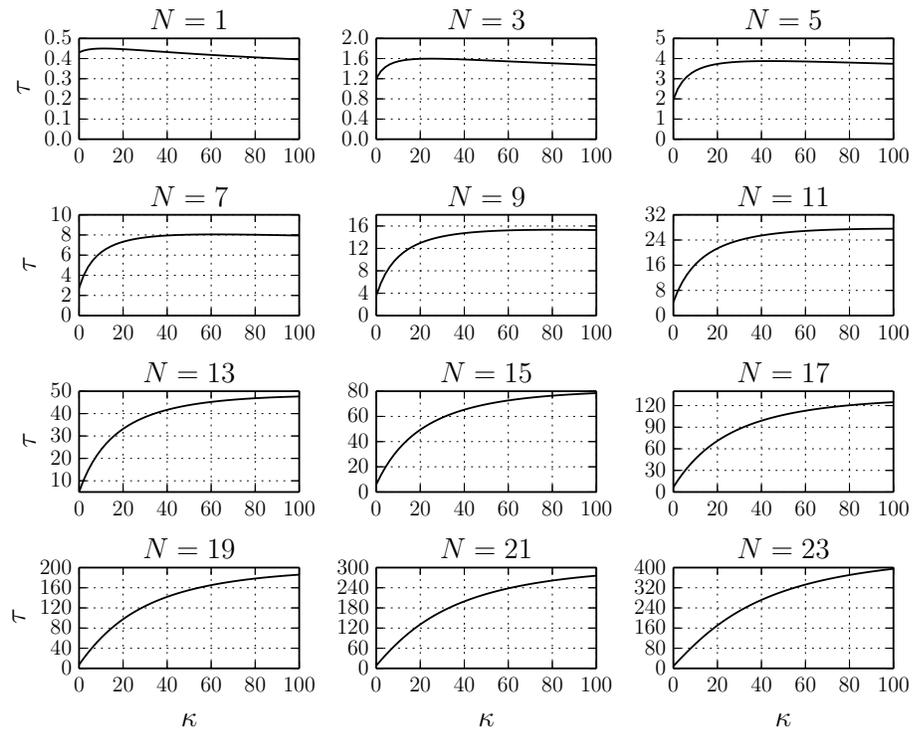}
\caption{\label{multitude}{\small{}$N$ is the number of catalyst particles in each thin zone layer.  For large
values of $N$ the system begins to  approach  the behavior of  a  two thin-zone network system
for which $\tau$ is  linear   in $\kappa$ with positive coefficients. }}
\end{centering}
\end{figure}

We first study the case $N=1$, that is, the system with two particles placed at different positions along the axis of the
cylinder. 
The fist graph (top left) of Figure \ref{multitude}
makes it clear that in this case
Formula \ref{twonodeformula}
does not represent well the two-particle system. In fact, if Formula \ref{twonodeformula} held, the experimental
$\tau(k)$, defined by \ref{tauexp}, would obey a  linear relation in $\kappa$ with positive coefficients, but this is clearly
not what we see in the graphs of Figure \ref{multitude}, when $N=1$. Therefore, one must look
for more complicated network approximations. (We do not do this here.)
On the other hand the dependence of conversion $\alpha_\kappa$ on $\kappa$
shown in Figure \ref{twozonealpha} is
qualitatively similar to that of 
the two thin-zone system.

What is most interesting about $\alpha_\kappa$ shown in Figure \ref{twozonealpha}
is that conversion attains a local minimum when the two particles are
at about the same place. Maximum $\alpha_\kappa$ occurs, naturally enough,
when the second particle is placed very close to the point of injection.
As the distance  of the second particle from the injection point increases past
the position of the first particle  conversion initially increases, then
reaches a maximum and then decreases as the second particle approaches the
reactor exit. Just as we have seen for the multi thin-zone network model,
this effect only happens for relatively large values of $\kappa$.  (Of course, for the network model
$\alpha_\infty=1$, whereas here the hitting probability must clearly  be less than $1$.)

Finally, we consider the effect of increasing the number $2N$ of particles in the experiment
associated to configuration (vi) of Figure \ref{six}. The question
is whether we can recover the two thin-zone network behavior by making 
the particles fill up the two zones more and more densely. 

More precisely,
the question is whether $\tau$ approaches for large $N$ the linear dependence on
$\kappa$ expected on the basis of the linear chain network  model with two active nodes. This is
precisely what the graph of Figure \ref{multitude} seems to suggest.
  As $N$ increases, the graph becomes more and more
``straight'' and seems to approach the straight line expected on the basis of the
network model.  The quality of this approximation depends on $\kappa$:   
for $\kappa<20$  and $N\geq 20$  the two-particle linear-chain network is a fairly good model for describing the behavior of  $\tau$. As  $\kappa$ increases, the nonlinearity of $\tau$
becomes more pronounced.

 The experiments of this subsection  raise the question of how the multiple thin zones system of
 Figure \ref{thinzones}, which is essentially $1$-dimensional, and the multiple particle $3$-dimensional systems are related. 
 In some respects they are very different as we see above in the behavior of $\tau$. On the other hand, the behavior of conversion itself
 is  qualitatively similar. This topic requires further analysis but it is not within the scope of this paper.

\section{Conclusions}
This paper investigates  a class of reaction-diffusion systems for  first order reaction $A\rightarrow B$
  in the presence of catalytic particles, where $A$ and $B$ are gas species. 
A pulse of $A$ is injected into the reactor at a given moment and the mixture of unreacted $A$ and
product $B$ is released at the reactor exit after diffusing (under ordinary Fickian diffusion with
constant diffusivity)  through the reactor bed.
We are interested in the ratio $\alpha$ of the number of $B$ molecules over the total number of molecules in
the gas outflow, which we call the  {\em conversion probability}. In particular, we are
interested in how $\alpha$ depends on the reaction rate constant as well as on the
geometric parameters that define reactor configuration, such as number, shapes and
distribution of the solid catalyst material.

  The main results of the paper are as follows:
  \begin{itemize}
\item   We have  described an effective and straightforward method based on the theory developed
  in \cite{matt1} for determining $\alpha$  by solving a time-independent boundary value problem for Laplace's equation
  that does not require solving first the diffusion equation for the concentrations of $A$ and $B$.
Using this method we have undertaken a systematic study of a class of reactor configurations that can 
  be investigated 
  experimentally in so-called TAP-systems. 
 \item We obtained $\alpha$  analytically  for a class of network model systems 
 for which the exact dependence of $\alpha$ on the reaction rate constant and other
 parameters of the system can be found. For general network models containing a single catalyst node,
 $\alpha$ takes on the expression
 $$\alpha_\kappa=\alpha_\infty\frac{\tau \kappa}{1+\tau\kappa} $$
 where $\kappa$ is the reaction constant and $\tau$ is a function of geometric parameters independent of $\kappa$.
 Here $\alpha_\infty$ is the  probability that a gas molecule will hit the catalyst. 
\item  We have tested the degree to which the formulas for $\alpha$ obtained for the network models serve 
 as useful approximations for the behavior of the more realistic $3$-dimensional systems.
 We have found that for a variety of reactor configurations consisting of a cylindrical reactor with
 a single, relatively small,  catalyst particle, the single catalyst node network model provides a resonably
 good approximation for dependence of $\alpha_\kappa$ on $\kappa$. 
 \item  For a single active site it is shown that the site activity can be significantly  enhanced  by modifying the architecture 
 of the network of channels in the vicinity of the particle. This network determines  the  local diffusional process and influences residence
 time at the particle location. 
 \item In the linear-chain multi-particle network,   it was shown that for low catalytic activity maximal conversion is obtained by placing 
 all particles close together near to the point of gas injection and for large catalytic activity the optimal configuration is the one in which
 the particles are equidistant. 
 \item In a series of numerical experiments, we investigated the influence  on conversion of many factors:  particle position 
 along the reactor axis or along the radius,  particle parameters (radius and length of the cylindrical particles), and distance between 
 two particles at the center of the reactor. Of these factors, it was shown that longitudinal position of the particle is the most significant.
 \item Finally, we investigated the following question: in a cylindrical reactor in which two groups of particles are placed  at
 two cross-sections, roughly evenly spaced in each group,  how many particles the two groups should have so
 that the system behaves  like a thin-zone system? More precisely, when is the expression for conversion
 obtained for the linear-chain, two-particle network a good model for the $3$-dimensional system with two groups of $N$ particles?
 We show that the answer depends on the catalytic activity, $\kappa$:  the larger the value of $\kappa$ the greater $N$ is needed to
 reproduce the behavior of the thin-zone system as far as the equation for computing conversion is concerned. 
  \end{itemize}

\end{document}